\documentclass[review,3p,twocolumn]{elsarticle}

\usepackage[hidelinks]{hyperref}

\usepackage{booktabs}       
\usepackage{pgfplots}       
\usepackage{pgfplotstable}  
\usepackage{tikz}           
\usetikzlibrary{patterns}
\usepackage{subcaption}

\makeatletter
\def\ps@pprintTitle{%
   \let\@oddhead\@empty
   \let\@evenhead\@empty
   \let\@oddfoot\@empty
   \let\@evenfoot\@oddfoot
}
\makeatother

\usepackage{listings}
\definecolor{strings_color}{rgb}{0.49,0.49,0.49}
\definecolor{comments_color}{rgb}{0.49,0.49,0.49}
\definecolor{border_color}{rgb}{0.81,0.81,0.81}
\usepackage{caption}
\usepackage{chngcntr} 

\journal{Arxiv.org}

\begin{document}

\begin{frontmatter}

\title{Exploring the Vision Processing Unit as Co-processor for Inference}





\author[kth]{Sergio~Rivas-Gomez}
\ead{sergiorg@kth.se}
\author[bsc]{Antonio~J.~Pe{\~n}a}
\ead{antonio.pena@bsc.es}
\author[int]{David~Moloney}
\ead{david.moloney@intel.com}
\author[kth]{Erwin~Laure}
\ead{erwinl@kth.se}
\author[kth]{Stefano~Markidis}
\ead{markidis@kth.se}
\address[kth]{KTH Royal Institute of Technology}
\address[bsc]{Barcelona Supercomputing Center (BSC)}
\address[int]{Intel Ireland Ltd.}

\begin{abstract}
The success of the exascale supercomputer is largely debated to remain dependent on novel breakthroughs in technology that effectively reduce the power consumption and thermal dissipation requirements. In this work, we consider the integration of co-processors in high-performance computing (HPC) to enable low-power, seamless computation offloading of certain operations. In particular, we explore the so-called Vision Processing Unit (VPU), a highly-parallel vector processor with a power envelope of less than 1W. We evaluate this chip during inference using a pre-trained GoogLeNet convolutional network model and a large image dataset from the ImageNet ILSVRC challenge. Preliminary results indicate that a multi-VPU configuration provides similar performance compared to reference CPU and GPU implementations, while reducing the thermal-design power (TDP) up to 8$\times$ in comparison.
\end{abstract}

\begin{keyword}
Vision Processing Unit\sep High-Performance Computing\sep Machine Learning
\end{keyword}

\end{frontmatter}

\section{Introduction}
\label{1_Introduction}

The recent advances in deep learning and convolutional networks, have dramatically influenced the role of machine learning on a wide-range of scientific applications~\cite{brink2013using, thompson2015machine}. This fact has been motivated by an increase in object classification and detection accuracy~\cite{szegedy2015going, long2015fully}, alongside with better tools for data mining that allow us to \mbox{understand large} datasets of unstructured information~\cite{wu2014data, witten2016data}. The inference error rate of machine learning algorithms has become remarkably low as well, reaching a state where the capacity of humans has been already surpassed in certain scenarios~\cite{ioffe2015batch}.

As a consequence, there is an existing trend
that proposes the integration of data-centric models on HPC that combines specialized hardware with the aim of fulfilling this need~\cite{sage2016whitepaper}. Upcoming major supercomputers are expected to feature new hardware architectures that provide high-performance 16-bit / 32-bit mixed arithmetic support for machine learning~\cite{top5002017oakridge}, both during training and inference.
In addition, innovation at software level is also observed with the appearance of novel data formats that use tensors with a shared exponent~\cite{courbariaux2014training, koster2017flexpoint}, maximizing the dynamic range of the traditional 16-bit floating point data format. These breakthroughs provide multiple advantages in terms of performance and power consumption. Specifically, some of the aforementioned architectural changes are expected to increase the performance 5--10$\times$ in comparison with current large-scale HPC clusters, using just twice the power~\cite{schneider2018us}. Hence, it will be of paramount importance for the success of the exascale supercomputer that we consider the embracement of these developments in the near-term future.

In this work, we set the initial steps towards the integration of low-power co-processors on HPC. In particular, we analyze the so-called \emph{Vision Processing Unit} (VPU). This type of processor emerges as a category of chips that aim to provide ultra-low power capabilities, without compromising performance. For this purpose, we explore the possibilities of the Movidius Myriad 2 VPU~\cite{moloney2014myriad, barry2015always} during inference in convolutional networks, over a large image dataset from the ImageNet ILSVRC 2012 challenge~\cite{ILSVRC15}. In our evaluations, we use a pre-trained network from the Berkeley Vision and Learning Center (BVLC), which follows the \emph{GoogLeNet} work by Szegedy et al.~\cite{szegedy2015going}. Preliminary results indicate that a combination of several of these chips can potentially provide equivalent performance compared to a reference CPU and GPU implementation, while reducing the thermal-design power (TDP) up to 8$\times$. The observed throughput, measured as number of inferences per Watt, is over 3$\times$ higher in comparison. The estimated top-1 error rate is 32\% on average, with a confidence error difference of 0.5\%. This is despite the differences in arithmetic precision (i.e., FP16).

The contributions of the work are the following:
    
\begin{itemize}
\item We provide a comprehensive technical overview of the Myriad 2 VPU in the context of the Intel Neural Compute Stick (NCS) platform~\cite{intel2017ncs}. 
\item We design and implement a small inference framework based on Caffe~\cite{jia2014caffe} and the Neural Compute API~\cite{intel2017ncsdk} to support our experiments on the VPU. 
\item We illustrate that VPUs feature an excellent ratio between throughput and power consumption compared to reference CPU and GPU implementations, including in multi-VPU configurations.
\item We compare the top-1 error rate~\cite{szegedy2015going} with a reference CPU implementation to understand the implications of using FP16 on the VPU.
\end{itemize}

The paper is organized as follows. We provide a high-level overview of the VPU in \autoref{2_Methods}. We describe the implementation considerations of a small inference framework in \autoref{2_1_Methods}. The experimental setup and performance evaluation is presented in \autoref{3_Results}. We extend the discussion of the results and provide further insights in \autoref{4_Discussion}. Related work is reported in \autoref{5_RelatedWork}. A summary of our conclusions and future work is outlined in \autoref{6_Conclusion}.

\section{Background}
\label{2_Methods}

The emergence of machine learning and data-centric applications on HPC poses several constraints on general-purpose processors, mainly due to the irregularity of the memory accesses that they feature~\cite{jang2011exploiting, yu2015imp}. These accesses have reduced temporal or spatial locality, incurring in long memory stalls and large bandwidth requirements. As a side effect, the power consumption and thermal dissipation requirements considerably increase as well~\cite{rhu2013locality}. Thus, during the last decade, scientists have experimented with the integration of novel algorithms that perform dynamic, in-memory data rearrangements of irregular structures~\cite{lloyd2015memory, hsieh2016accelerating}. The aim is to overcome (or partially hide) some of the aforementioned limitations.

\begin{figure}
    \vspace{0.0921cm}
    \begin{center}
        \includegraphics[width=0.8921\columnwidth]{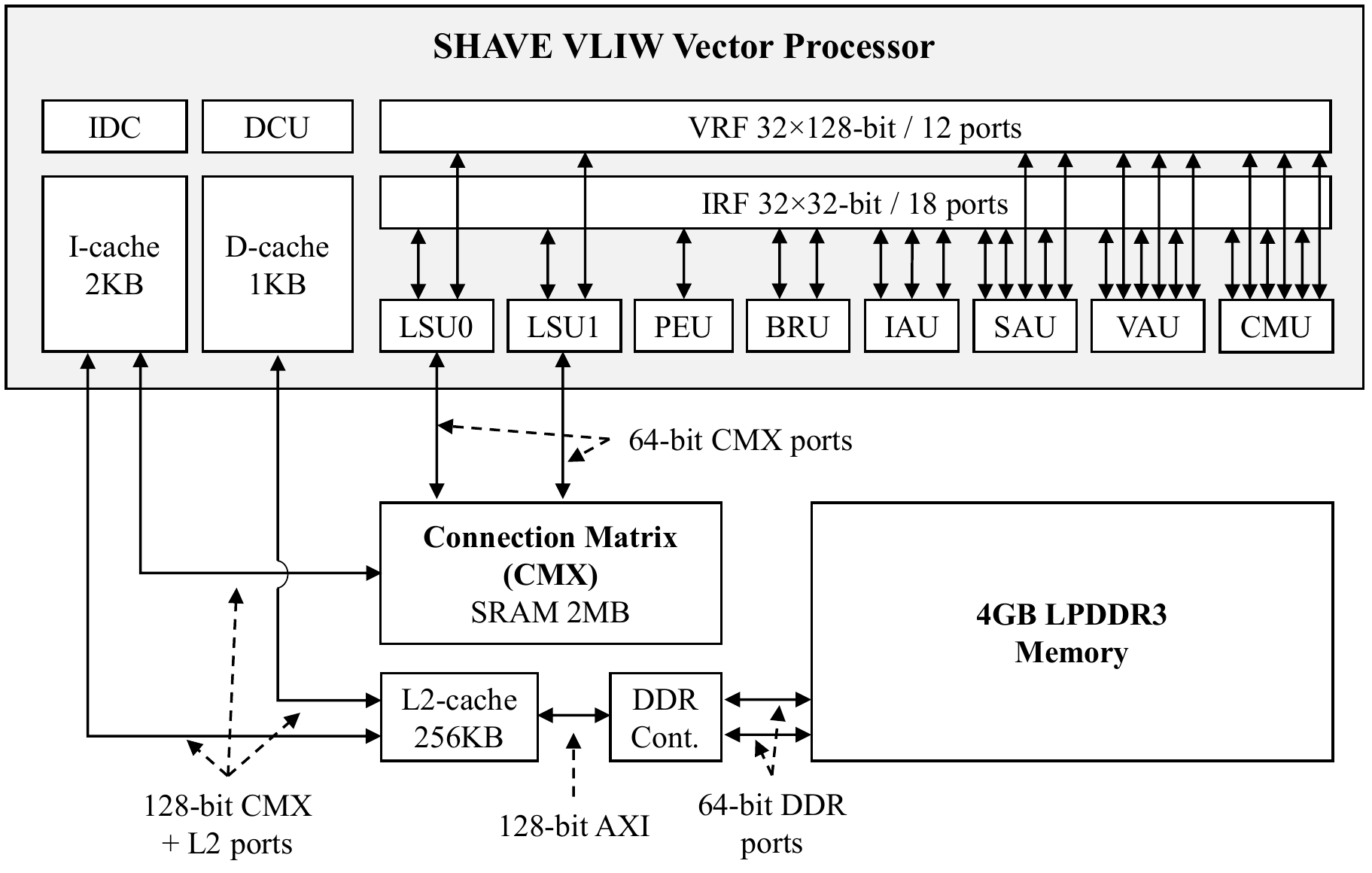}
        \caption{High-level representation of one of the SHAVE vector processors featured on the Myriad 2 VPU~\cite{barry2015always}. The Connection Matrix (CMX) enables seamless interaction between the vector processors and other hardware components.}
        \label{fig:shave}
    \end{center}
\end{figure}

\begin{figure}
    \begin{center}
        \includegraphics[width=0.8921\columnwidth]{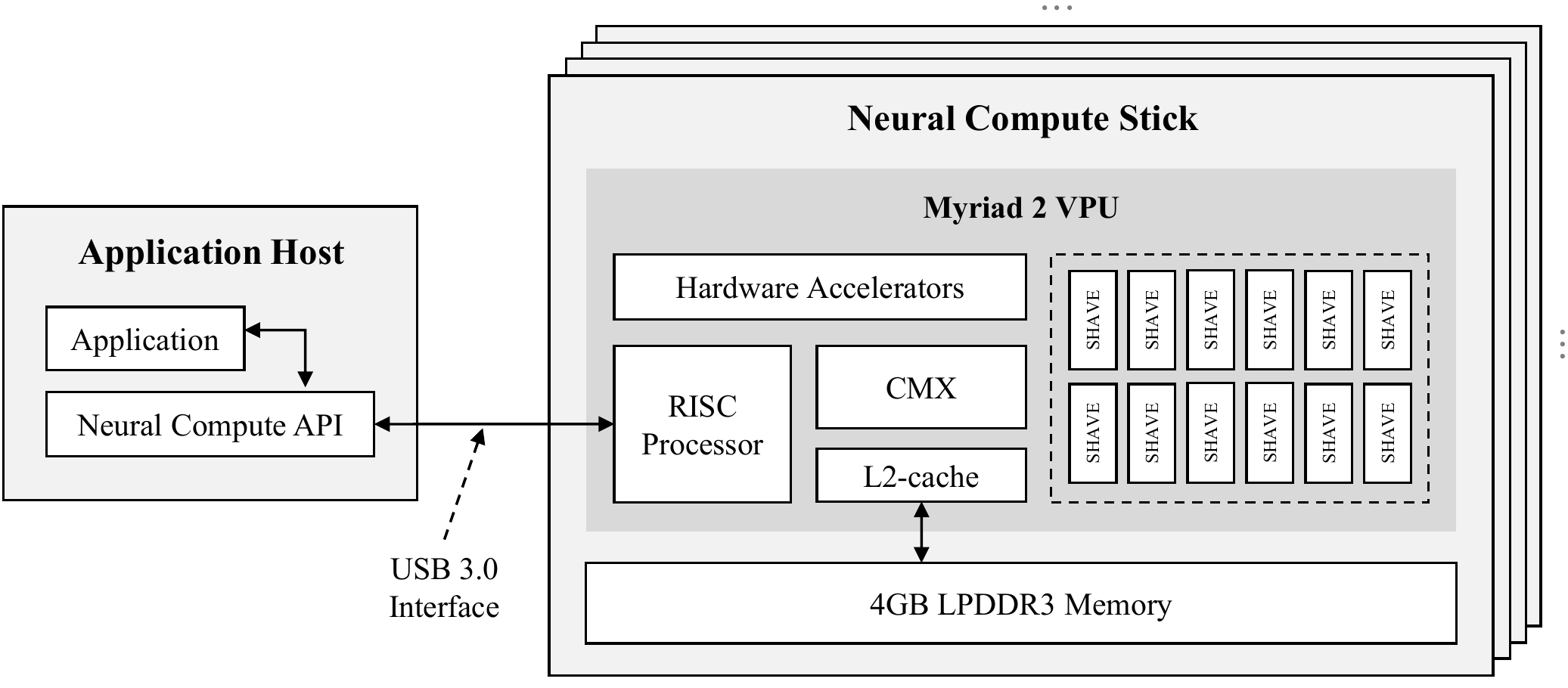}
        \caption{Approximate implementation of the Myriad 2 VPU used within the Neural Compute Stick (NCS) platform~\cite{intel2017ncs}. The Neural Compute API allows us to coordinate the execution on the VPU of one or more NCS devices~\cite{intel2017ncsdk}.}
        \label{fig:ncs}
    \end{center}
\end{figure}

Nonetheless, the inherent complexity of such techniques, coupled with the adoption of the ``CPU + Accelerator'' model to enhance the performance of scientific applications~\cite{nickolls2010gpu}, makes programming general-purpose processors another key-factor to consider. In addition, transferring data among these different hardware layers can also become costly~\cite{gregg2011data}.
As a consequence, the industry is shifting towards designing processors where cost, power, and thermal dissipation are key concerns~\cite{barry2015always}. Specialized co-processors have recently emerged with the purpose of reducing the power envelope constraints, while improving the overall performance on scenarios such as machine learning~\cite{ionica2015movidius}. In this regard, we observe that other scientific fields can benefit from this trend by adopting part of these technologies. In fact, energy consumption in HPC is considered one of the main limiting factors towards the exascale supercomputer~\cite{dongarra2011international}.

In this section, we briefly describe the most relevant technical aspects of the Movidius Myriad 2 VPU~\cite{moloney2014myriad, barry2015always}, in the context of the Intel Neural Compute Stick (NCS) platform~\cite{intel2017ncs}. Our goal is to understand how this type of low-power co-processors could potentially be integrated for computation offloading on HPC. 

\subsection{Vision Processing Unit}

The Myriad 2 VPU is designed as a 28-nm co-processor that provides high-performance tensor acceleration. The chip dissipates less than 1W~\cite{moloney2014myriad}. High-level APIs allow application programmers to easily take advantage of its features and, thus, enhance programming productivity. In addition, the software-controlled memory subsystem enables fine-grained control on different workloads, if required. The term ``vision'' is employed due to its original purpose, which was meant to accelerate computer vision applications on the ``edge''~\cite{moloney20111tops}.

The architecture of this chip is inspired by Agarwal's observation, which states that beyond a certain frequency limit for any particular design and target process technology, the cost is quadratic in power for linear increases in operating frequency~\cite{barry2015always}. Following this statement, the Myriad 2 VPU is designed featuring 12 highly-parallelizable vector processors, named Streaming Hybrid Architecture Vector Engines (SHAVE). Each SHAVE processor contains wide register files and several functional units. These are controlled by Variable-Length Long Instruction Word (VLLIW) packets. Hence, enabling seamless SIMD operations on the chip. The nominal frequency is 600MHz.

\begin{figure}
    \vspace{0.0921cm}
    \begin{center}
        \includegraphics[width=0.8921\columnwidth]{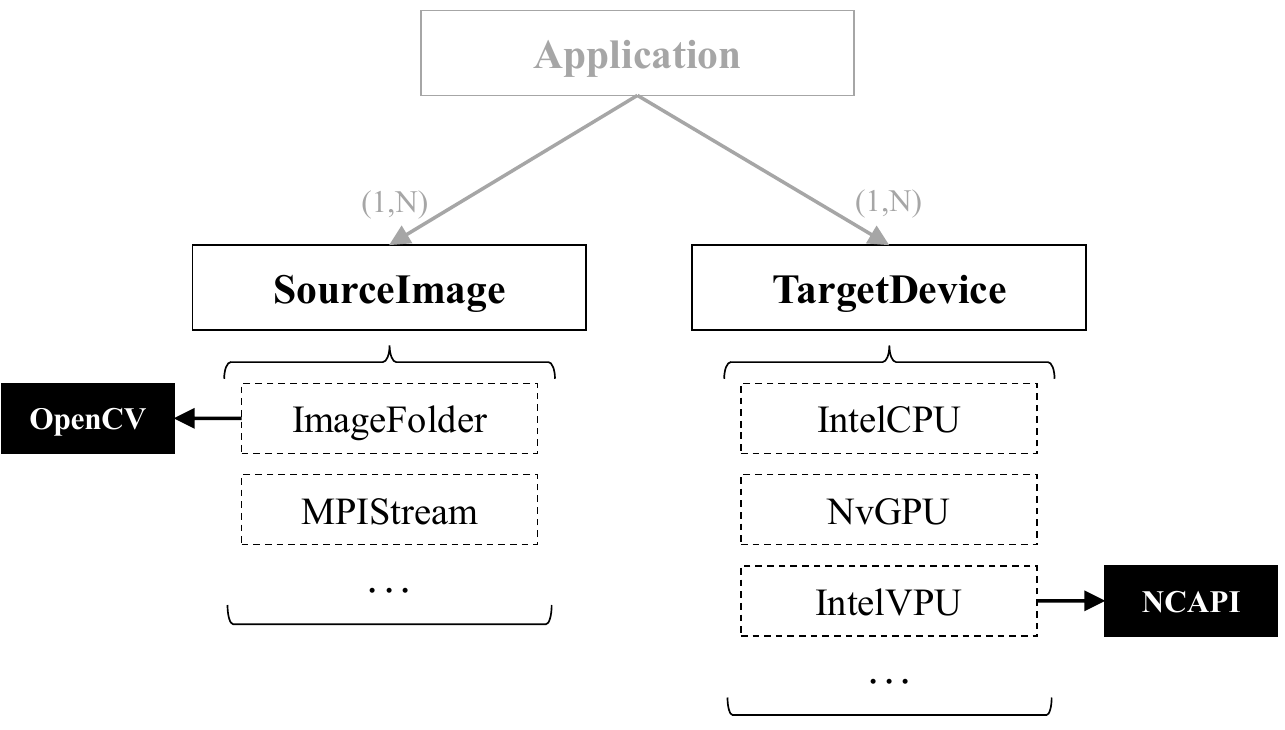}
        \caption{Class diagram specification of the NCSw framework. The simple modular design allows us to provide implementations for new kind of devices (e.g., FPGA).}
        \label{fig:class_diagram}
    \end{center}
\end{figure}

\begin{figure}
\begin{lstlisting}
...
// Load the graph with the input image
mvncLoadTensor(graph, img, size, NULL);

/************************************
 * Perform overlapping computations *
 ************************************/

// Retrieve the result from the NCS
mvncGetResult(graph, (half **)&result,
              &result_size, &userParam);
...
\end{lstlisting}
\vspace{-0.21cm}
\captionof{lstlisting}{Source code example in C that illustrates how to conduct inference on the NCS using the NCAPI~\cite{intel2017ncsdk}.\label{exampleCode}}
\end{figure}

\autoref{fig:shave} illustrates a high-level diagram of one of the SHAVE processors and the interactions with other components of the Myriad 2 VPU. The main vector register file (VRF) has 128-bit $\times$ 32 entries and 12 ports. A general register file (IRF) is also available with \mbox{32-bit $\times$ 32 entries} and 18 ports. Among the functional units of each SHAVE processor,
we highlight the 128-bit Vector Arithmetic Unit (VAU), the 128-bit Compare-and-Move Unit (CMU), the 32-bit Scalar Arithmetic Unit (SAU), and the 32-bit Integer Arithmetic Unit (IAU). The chip supports 8, 16, 32, and partially 64-bit integer operations, as well as native FP16 and FP32 arithmetic\footnote{The maximum theoretical performance claimed by the manufacturer is 1000 Gflops using FP16 arithmetic~\cite{barry2015always}.}. Each of these functional units can be operated independently through the VVLIW instruction packets. In addition, two 64-bit Load-and-Store Units (LSU) enable data transferring among the SHAVE processors through a shared, multi-ported 2MB memory block, named Connection Matrix (CMX). The CMX features 16 blocks of 128KB, comprising four 32KB RAM instances organized as 4096 words of 64-bits each, independently arbitrated. The variant used in our tests (MA2450) features a global stacked memory of 4GB LPDDR3. The memory fabric of the Myriad 2 VPU is designed for low-latency by endorsing data locality. It is also mostly software-controlled for flexibility purposes, as previously stated. This allows the VPU to support different kinds of application workloads.

Alongside the SHAVE vector processors, the chip features a Streaming Image Processing Pipeline (SIPP), which contains fully programmable hardware-accelerated kernels of common image processing operations~\cite{moloney2014myriad}. For instance, some of the kernels include tone-mapping, Harris Corner detector, Histogram of Oriented Gradients (HoG) edge operator, luminance / chrominance denoising, and others. The typical configuration for the kernels is 5$\times$5 per target output pixel. Each hardware-accelerated kernel is connected to the CMX memory block using a crossbar. A local controller on each SIPP filter manages the read / writeback of the results to the CMX. Thus, combining operations on the SHAVE vector processors and the hardware-accelerated kernels is feasible. The filters can output completely computed pixels individually per cycle.

\subsection{Neural Compute Stick Platform}

The Intel Neural Compute Stick (NCS) platform~\cite{intel2017ncs} is a System-on-Chip (SoC) implementation of the Myriad 2 VPU. A high-level overview of the device is illustrated in~\autoref{fig:ncs}. The diagram depicts the approximate implementation used in the NCS platform (variant MA2450). The NCS employs a total of 20 power islands, including one for each of the 12 integrated SHAVE processors. This is critical to effectively manage the power consumption of the SoC. Two RISC processors manage the communication with the host and the execution on the VPU (i.e., runtime scheduler). They are also in charge of the peripherals in other implementations (e.g., MIPI D-PHY) and running a Unix-based real-time OS (RTOS). In the diagram, applications communicate with the VPU using a USB 3.0 interface and the so-called Neural Compute API (NCAPI)~\cite{intel2017ncsdk}. The main purpose of this API is to enable the deployment of convolutional networks for inference on the NCS\footnote{Training these networks, however, is accomplished outside the scope of the NCS using the regular Caffe~\cite{jia2014caffe} or Tensorflow~\cite{abadi2016tensorflow} frameworks (i.e., the device is only used for inference).}. When the NCAPI initializes and opens a device, a firmware is loaded onto the NCS. At this point, the device is ready to accept the network graph files and execute commands to conduct inference on the VPU.

The NCAPI comprehends a set of operations that allow applications to connect to the NCS, deploy a pre-trained convolutional network model, obtain performance metrics per layer, and more. The programming interface is available in C/C++ and Python. For instance, in order to perform inference on the device, the API follows a set of operations that resemble the MPI non-blocking interface~\cite{gropp2014using}. In this case, instead of having a single, blocking ``\texttt{inference()}'' function, the step is divided in two separate operations. First, a \emph{load} operation transfers the input and prepares the NCS for execution. Thereafter, a \emph{wait} operation blocks the process on the host until the execution on the NCS has finished. Hence, this model enables the design of decoupled strategies that overlap computations while inference has been offloaded to the NCS. 

Listing~\ref*{exampleCode} provides a source code example in C where the NCAPI is utilized to perform inference on the VPU. Error-checking is excluded for illustration purposes. In this example, the \texttt{mvncLoadTensor()} transfers a certain input image to the NCS device and loads the pre-compiled graph for execution. This will automatically coordinate the data transfer with one of the RISC processors into the NCS. It will also immediately queue the execution of the graph on the SHAVE processors through the runtime scheduler. The operation will return as soon as the data is transferred and the execution is scheduled, without blocking the host process. At this point, the application is able to overlap additional computations while the inference has been offloaded to the NCS (e.g., decode the next frame). Multi-device is also supported, meaning that we could easily offload more inference operations to other devices. When the result is required, a call to \texttt{mvncGetResult()} will guarantee that the host process is blocked until the inference has finished and the result is ready. The output result is a list of labels with the correspondent confidence.

Note that fine-grained general-purpose computing using C/C++ is also possible through the Movidus Development Kit (MDK)~\cite{ionica2015movidius, movidius2016mdk}. The MDK enables OpenCL support and provides several optimized libraries designed for the Myriad 2 VPU chip (e.g., LAMA, a linear algebra library). Tools for debugging and profiling are also available. We consider exploring the possibilities of the MDK for general-purpose computing in future work.

\section{Inference Framework}
\label{2_1_Methods}

We design and implement a very simple inference framework using C/C++ that supports diverse types of target devices. The framework, named Neural Compute Stick Wrapper (NCSw), is mostly based on the use of Caffe~\cite{jia2014caffe} in the context of the NCS platform. In addition, we integrate the specific Caffe project forks optimized for Intel processors and NVIDIA graphics cards to conduct our experiments. This allows us to compare the inference performance with the VPU chip. The source code is available on a public Git repository\footnote{\url{https://github.com/sergiorg-kth/ncs-wrapper}}.

\begin{figure}
    \vspace{0.0921cm}
    \begin{center}
        \includegraphics[width=0.921\columnwidth]{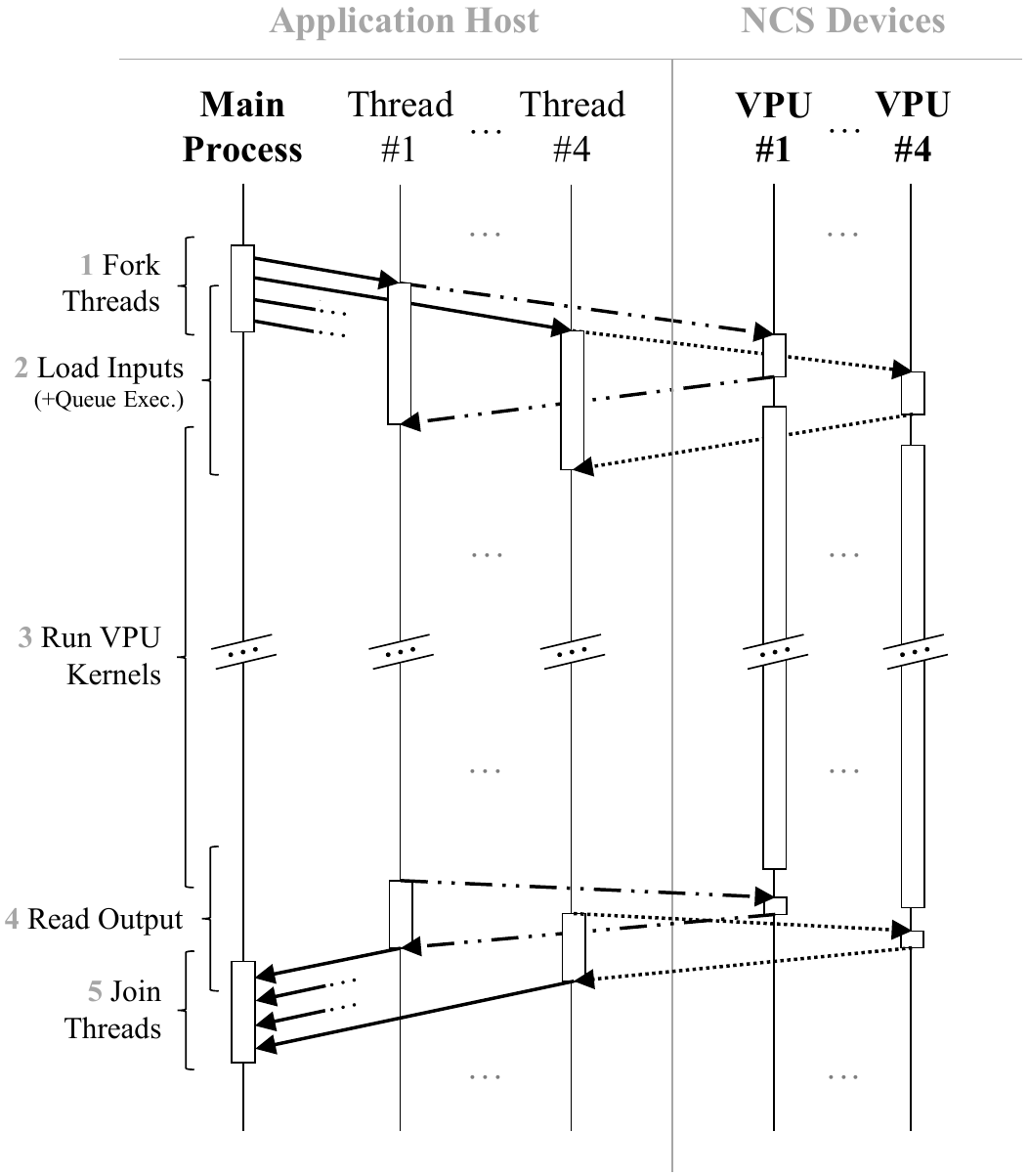}
        \vspace{-0.21cm}
        \caption{Execution timeline of the parallel, multi-VPU implementation of NCSw. In this example, the host process offloads the execution to four threads, one per NCS device.}
        \label{fig:multiVPU}
    \end{center}
\end{figure}

The NCSw framework is divided in several abstract classes that represent the \emph{source} of the input datasets \mbox{and the} \emph{target} (or where) to conduct inference (\autoref{fig:class_diagram}). The aim is to provide an easy-to-use implementation that could enable the integration of new kinds of input sources (e.g., MPI streams~\cite{peng2015data}) or target devices (e.g., FPGA) in the future. The VPU implementation is based on the use of the NCAPI. We use OpenEXR~\cite{kainz2009technical} half-precision class for converting the pixel data from FP32 to FP16 (i.e., the compatible format~\cite{barry2015always}).

\begin{figure}
    \vspace{0.21cm}
    \begin{center}
        \includegraphics[width=1.0\columnwidth, trim=1.21cm 21.96cm 0.0cm 16.81cm, clip]{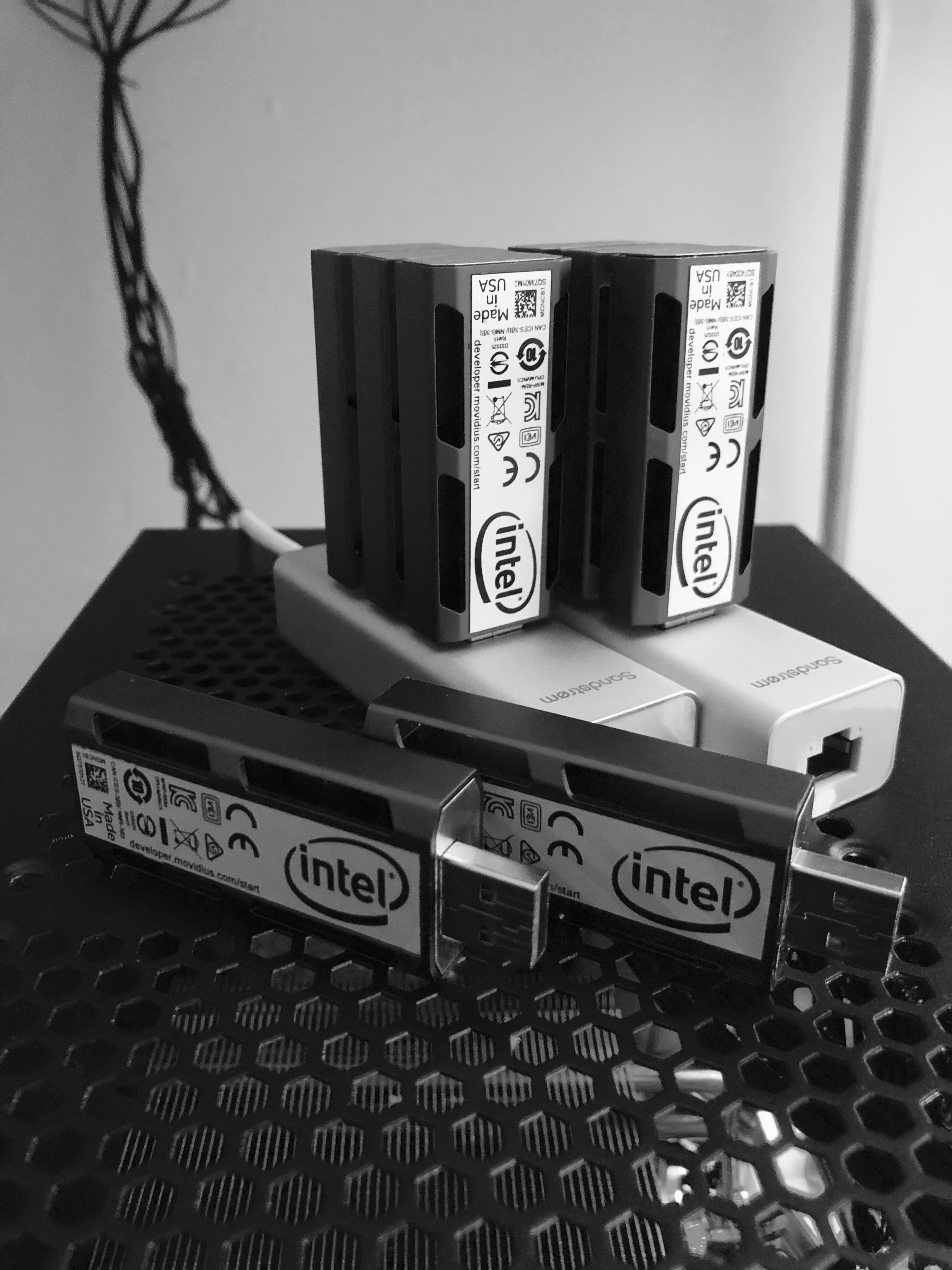}
        \caption{Our testbed contains 8 different NCS devices, where 6 devices are connected using two USB 3.0 HUBs and 2 devices are connected using the ports of the motherboard.}
        \label{fig:testbed}
    \end{center}
\end{figure}

Batch-processing is supported by defining a parallel, multi-VPU implementation. This approach differs from the traditional Caffe batched execution, which resizes the input blob layer of the convolutional network to achieve better data communication throughput (e.g., on GPUs). In this case, we schedule simultaneous inferences using the same graph on multiple NCS devices. The main host process is responsible for connecting to each device and offloading the execution. By default, if the NCSw framework is compiled with OpenMP support, the multi-VPU implementation will become multi-threaded. Hence, the host process will spawn multiple threads to handle the execution on each NCS device available. The threads will concurrently transfer the source input and retrieve the output, thus, effectively overlapping the communication with the RISC processor on the SoC and maximizing the bandwidth utilization.

\autoref{fig:multiVPU} illustrates an example timeline for the parallel, multi-VPU implementation using four different NCS devices. Here, the host process begins by spawning one thread per VPU. These threads will then load different inputs into the global LPDDR3 memory of each NCS. This fact will guarantee that, while the next input is being loaded on the succeeding device, the runtime scheduler in the preceding device has started the execution on the SHAVE processors and SIPP hardware-accelerated filters. Thereafter, the results are retrieved in the queueing order to guarantee an overlap with the rest of the NCS devices. We follow a simple static scheduling (i.e., round-robin).

Applications can decide whether to use one or more VPUs simultaneously, or to define groups of the same target type. In other words, different sources can be easily connected to the same or multiple targets. Therefore, some applications might choose to run a specific subset of inputs on a GPU, and at the same time another subset on two different groups that connect to several VPUs using the described approach.

\section{Experimental Results}
\label{3_Results}

In this section, we analyze three implementations inside the NCSw framework that target a CPU, a GPU, and a multi-VPU configuration, respectively. We evaluate these implementations in terms of inference performance and confidence error. For this purpose, we use the Intel-optimized Caffe-MKL fork (v1.0.7) for Intel processors, the NVIDIA-optimized Caffe-cuDNN fork (v0.16.4) for NVIDIA graphic cards, and the Neural Compute SDK (v1.12.00.01) for the Myriad 2 VPU on the NCS. Thus, we aim to take advantage of each of these devices using reference implementations provided by the manufacturers.

The simulations are conducted in a workstation with two four-core Intel Xeon E5-2609v2 processors running at 2.5GHz. The workstation is equipped with a total of 72GB DRAM. The graphics card is a Quadro K4000, with 3GB of GDDR5 and 768 CUDA cores. The NVIDIA driver version is v384.81. The storage consists of two 4TB HDD (WDC WD4000F9YZ / non-RAID) and a 250GB SSD (Samsung 850 EVO). The OS is Ubuntu Server 16.04.1 LTS with Kernel 4.4.0-62-generic. The NCSw framework is compiled with {\ttfamily gcc} v5.4.0 and linked with OpenCV v2.4.9.1 to decode the input images. We compile Caffe with Intel MKL v2018.1.163. For the GPU version, we use CUDA v9.0, cuDNN v7.0.5.15-1, and NCCL v1.3.4-1.

Note that all the figures reflect the standard deviation of the samples as error bars. In addition, we omit from our results the decoding time per image, but account for the data transferring time from host to device. We also enable OpenMP to support multi-threading on the multi-VPU configuration with a maximum of 8 simultaneous NCS devices. 6 devices are connected using two USB 3.0 HUBs (Sandstr{\o}m 164903) and 2 devices are connected using directly the USB 3.0 ports of the motherboard (\autoref{fig:testbed}). Lastly, we use traditional Caffe batch-based processing on the CPU and GPU.

\subsection{Performance Evaluation}

With the purpose of evaluating the image classification performance of the three aforementioned implementations, we use one of the reference datasets from the ImageNet database~\cite{ILSVRC15}. This project is an on-going research effort that aims to provide researchers around the world with an easily accessible, large image database organized according to the WordNet hierarchy~\cite{miller1995wordnet}. Each meaningful concept in ImageNet is described by multiple word phrases (i.e., "synonym set" or "synset"), and contains on average 1000 images per definition.

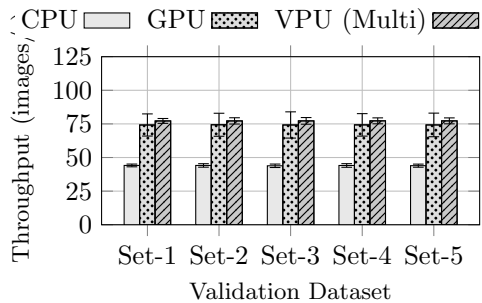
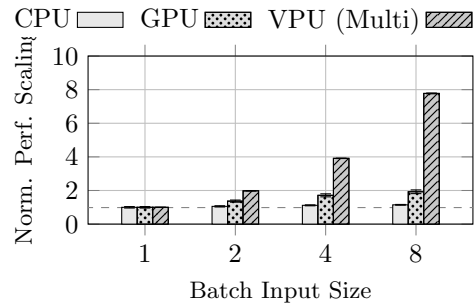
\begin{figure*}[t]
    \centering
    \begin{subfigure}[t]{0.4921\textwidth}
        \centering
        \hspace{-0.3921cm}
        \begin{tikzpicture}
            \begin{axis}[
                xlabel=Validation Dataset,
                ylabel=Throughput (images/s),
                symbolic x coords={Set-1,Set-2,Set-3,Set-4,Set-5},
                xtick=data,
                ytick={0,25,50,75,100,125},
                ymin=0,
                ymax=125,
                scaled y ticks = false,
                y tick label style={/pgf/number format/fixed, /pgf/number format/1000 sep = },
                ylabel style={at={(0.03921,0.521)}, style={font=\small}},
                xlabel style={at={(0.5,-0.05)}, style={font=\small}},
                enlarge x limits=0.16,
                grid=major,
                legend style={legend columns=-1,at={(1.0,1.321)},anchor=north east},
                legend style={/tikz/every even column/.append style={column sep=4.21}},
                legend cell align=right,
                legend plot pos=right,
                legend style={draw=none, fill=white, inner xsep=0, inner ysep=0},
                ybar=0pt,
                bar width=5.81pt,
                width=0.81\textwidth,
                height=3.81cm,
                area legend
            ]
            
            \addplot [fill=black!9] plot[error bars/.cd, y dir=both, y explicit] table [y error minus=min, y error plus=max] {
                x       y               min             max
                Set-1	44.1165822597	1.0224807568	1.0224807568
                Set-2	44.1111871214	1.2952748714	1.2952748714
                Set-3	43.8508830332	1.2870644736	1.2870644736
                Set-4	44.0888983415	1.350238924	1.350238924
                Set-5	43.942247539	1.1540176617	1.1540176617
            }; \addlegendentry{CPU}
            
            \addplot [fill=black!13, postaction={pattern=crosshatch dots}] plot[error bars/.cd, y dir=both, y explicit] table [y error minus=min, y error plus=max] {
                x       y               min             max
                Set-1	74.2016707174	8.2398902749	8.2398902749
                Set-2	74.3666317147	8.5571214525	8.5571214525
                Set-3	74.2151825617	9.7580715979	9.7580715979
                Set-4	74.2774443412	8.3617278858	8.3617278858
                Set-5	74.237281989	8.7809008927	8.7809008927
            }; \addlegendentry{GPU}
            
            \addplot [fill=black!21, postaction={pattern=north east lines}] plot[error bars/.cd, y dir=both, y explicit] table [y error minus=min, y error plus=max] {
                x       y               min             max
                Set-1	77.2175736117	1.8085856308	1.8085856308
                Set-2	77.2434045791	2.3342252015	2.3342252015
                Set-3	77.2530976995	2.4743050815	2.4743050815
                Set-4	77.2806664561	2.2132393449	2.2132393449
                Set-5	77.2612658273	2.2117521899	2.2117521899
            }; \addlegendentry{VPU (Multi)}
            \end{axis}
        \end{tikzpicture}
        \caption{\textbf{Inference Performance} / 8$\times$Input (batch)} \label{fig:benchmarks_inference}
    \end{subfigure}
    \hfill
    \begin{subfigure}[t]{0.4921\textwidth}
        \centering
        \hspace{-0.2921cm}
        \begin{tikzpicture}
            \begin{axis}[
                xlabel=Batch Input Size,
                ylabel=Norm. Perf. Scaling,
                symbolic x coords={1,2,4,8},
                xtick=data,
                ytick={0,2.0,4.0,6.0,8.0,10.0},
                ymin=0,
                ymax=10.0,
                scaled y ticks = false,
                y tick label style={/pgf/number format/fixed, /pgf/number format/1000 sep = },
                ylabel style={at={(0.081,0.521)}, style={font=\small}},
                xlabel style={at={(0.5,-0.05)}, style={font=\small}},
                enlarge x limits=0.21,
                grid=major,
                legend style={legend columns=-1,at={(1.0,1.321)},anchor=north east},
                legend style={/tikz/every even column/.append style={column sep=4.21}},
                legend cell align=right,
                legend plot pos=right,
                legend style={draw=none, fill=white, inner xsep=0, inner ysep=0},
                ybar=0pt,
                bar width=5.81pt,
                width=0.821\textwidth,
                height=3.81cm,
                area legend
            ]
            
            \draw  (-1.21cm,0.21921cm) -- (5.81cm,0.21921cm) [dashed, draw=black!49];
            
            \addplot [fill=black!9] plot[error bars/.cd, y dir=both, y explicit] table [y error minus=min, y error plus=max] {
                x       y               min             max
                1	1	0.0441534296	0.0441534296
                2	1.0558745785	0.0376682925	0.0376682925
                4	1.1179011838	0.0293716359	0.0293716359
                8	1.1474006592	0.0231767899	0.0231767899
            }; \addlegendentry{CPU}
            
            \addplot [fill=black!13, postaction={pattern=crosshatch dots}] plot[error bars/.cd, y dir=both, y explicit] table [y error minus=min, y error plus=max] {
                x       y               min             max
                1	1	0.0447499517	0.0447499517
                2	1.367688891	0.0641090049	0.0641090049
                4	1.7125882367	0.0907529874	0.0907529874
                8	1.9250280417	0.1110472338	0.1110472338
            }; \addlegendentry{GPU}
            
            \addplot [fill=black!21, postaction={pattern=north east lines}] plot[error bars/.cd, y dir=both, y explicit] table [y error minus=min, y error plus=max] {
                x       y               min             max
                1	1	0.0034748992	0.0034748992
                2	1.9819247956	0.0043604443	0.0043604443
                4	3.9130582503	0.0119728724	0.0119728724
                8	7.7763129668	0.0234219433	0.0234219433
            }; \addlegendentry{VPU (Multi)}
            \end{axis}
        \end{tikzpicture}
        \caption{\textbf{Relative Inference Performance} per batch size} \label{fig:benchmarks_scaling}
    \end{subfigure}
    \caption{(\subref*{fig:benchmarks_inference}) Inference performance (per subset) of the ILSVRC 2012 Validation dataset using batch-mode on the CPU, GPU, and multi-VPU configurations. (\subref*{fig:benchmarks_scaling}) Normalized performance scaling per batch size relative to the baseline of a single input for each device type. Note that, in both figures, the number of active VPU chips is coupled with the input size (1-8).\label{fig:benchmarks}}
\end{figure*}

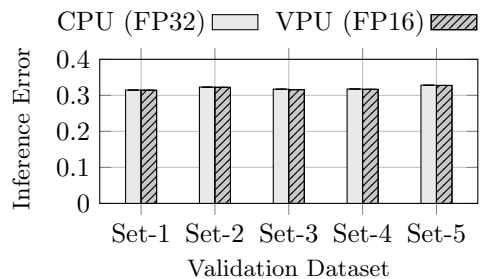
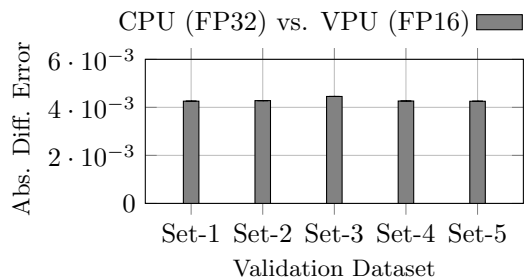
\begin{figure*}[t]
    \centering
    \begin{subfigure}[t]{0.4921\textwidth}
        \centering
        \hspace{-0.3921cm}
        \begin{tikzpicture}
            \begin{axis}[
                xlabel=Validation Dataset,
                ylabel=Inference Error,
                symbolic x coords={Set-1,Set-2,Set-3,Set-4,Set-5},
                xtick=data,
                ytick={0.0,0.1,0.2,0.3,0.4},
                ymin=0,
                ymax=0.4,
                scaled y ticks = false,
                y tick label style={/pgf/number format/fixed, /pgf/number format/1000 sep = },
                ylabel style={at={(0.03921,0.521)}, style={font=\small}},
                xlabel style={at={(0.5,-0.05)}, style={font=\small}},
                enlarge x limits=0.14,
                grid=major,
                legend style={legend columns=-1,at={(1.0,1.381)},anchor=north east},
                legend style={/tikz/every even column/.append style={column sep=4.21}},
                legend cell align=right,
                legend plot pos=right,
                legend style={draw=none, fill=white, inner xsep=0, inner ysep=0},
                ybar=0pt,
                bar width=5.81pt,
                width=0.81\textwidth,
                height=3.4921cm,
                area legend
            ]
            
            \addplot [fill=black!9] plot[error bars/.cd, y dir=both, y explicit] table [y error minus=min, y error plus=max] {
                x       y               min             max
                Set-1   0.3147 0 0
                Set-2   0.3228 0 0
                Set-3   0.3171 0 0
                Set-4   0.3176 0 0
                Set-5   0.3281 0 0
            }; \addlegendentry{CPU (FP32)}
            
            \addplot [fill=black!21, postaction={pattern=north east lines}] plot[error bars/.cd, y dir=both, y explicit] table [y error minus=min, y error plus=max] {
                x       y               min             max
                Set-1   0.3141 0 0
                Set-2   0.3221 0 0
                Set-3   0.3160 0 0
                Set-4   0.3166 0 0
                Set-5   0.3274 0 0
            }; \addlegendentry{VPU (FP16)}
            \end{axis}
        \end{tikzpicture}
        \caption{\textbf{Top-1 Inference Error} per subset} \label{fig:benchmarks_top1}
    \end{subfigure}
    \hfill
    \begin{subfigure}[t]{0.4921\textwidth}
        \centering
        \hspace{-1.21cm}
        \begin{tikzpicture}
            \begin{axis}[
                xlabel=Validation Dataset,
                ylabel=Abs. Diff. Error,
                symbolic x coords={Set-1,Set-2,Set-3,Set-4,Set-5},
                xtick=data,
                ytick={0.0,0.002,0.004,0.006},
                ymin=0,
                ymax=0.006,
                scaled y ticks = false,
                ylabel style={at={(-0.081,0.521)}, style={font=\small}},
                xlabel style={at={(0.5,-0.05)}, style={font=\small}},
                enlarge x limits=0.16,
                grid=major,
                legend style={legend columns=-1,at={(1.0,1.381)},anchor=north east},
                legend style={/tikz/every even column/.append style={column sep=4.21}},
                legend cell align=right,
                legend plot pos=right,
                legend style={draw=none, fill=white, inner xsep=0, inner ysep=0},
                ybar=0pt,
                bar width=5.81pt,
                width=0.81\textwidth,
                height=3.4921cm,
                area legend
            ]
            
            \addplot [fill=black!49] plot[error bars/.cd, y dir=both, y explicit] table [y error minus=min, y error plus=max] {
                x       y               min             max
                Set-1   0.0042577629 0 0
                Set-2   0.0042805113 0 0
                Set-3   0.0044536024 0 0
                Set-4   0.0042644211 0 0
                Set-5   0.0042540864 0 0
            }; \addlegendentry{CPU (FP32) vs. VPU (FP16)}
            \end{axis}
        \end{tikzpicture}
        \caption{\textbf{Confidence Difference} per subset} \label{fig:benchmarks_diff}
    \end{subfigure}
    \caption{(\subref*{fig:benchmarks_top1}) Top-1 prediction error (per subset) of the ILSVRC 2012 Validation dataset using the CPU (FP32) and VPU (FP16) implementations. (\subref*{fig:benchmarks_diff}) Absolute confidence difference error after filtering the top-1 miss-predictions between the CPU and VPU implementations.\label{fig:benchmarks_error}}
\end{figure*}

The success of ImageNet is largely due to the Large Scale Visual Recognition Challenge (ILSVRC). This challenge is a benchmark in object category classification and detection on hundreds of object categories and millions of images. Since its inception in 2010, ILSVRC has become the de-facto standard benchmark for large-scale object recognition~\cite{imagenet2017quartz}. The publicly released dataset contains a set of manually annotated training and test images.

In this regard, we use the \emph{Validation} dataset from the ILSVRC 2012 challenge\footnote{\url{http://image-net.org/challenges/LSVRC/2012}} to conduct our experiments. This dataset contains 50000 images in total. Each target device in our implementation uses the pre-trained BAIR GoogLeNet network\footnote{\url{http://dl.caffe.berkeleyvision.org/bvlc_googlenet.caffemodel}} from the Berkeley Vision and Learning Center (BVLC). This network is trained specifically for the ILSVRC 2012 challenge, as described by Szegedy et al.~\cite{szegedy2015going}. The input geometry of the network is 224x224. The mean values are retrieved directly from the ILSVRC 2012 training dataset. Finally, the Caffe engine is set to ``MKL2017'' for the CPU-based implementation.

Using a multi-VPU configuration, we determine that the Myriad 2 VPU provides a very well-balanced ratio between performance and power consumption. \autoref{fig:benchmarks_inference} reports the throughput in images per second (img{$\cdot$}s\textsuperscript{-1}) for the CPU, GPU, and multi-VPU configurations. We use batch-processing mode with 8 inputs to match the number of simultaneous VPUs available in our testbed (i.e., eight NCS devices). For evaluation purposes, we divide the complete validation dataset in groups of 10000 images, forming 5 subsets in total. From this figure, we can determine that the throughput using eight Myriad 2 VPU chips is approximately 77.2 img{$\cdot$}s\textsuperscript{-1} (12.9ms per inference). The optimized Caffe framework on the CPU is 40.7\% slower, with an average of 44.0 img{$\cdot$}s\textsuperscript{-1} (22.7ms per inference). However, the GPU-based implementation produces similar results, with a throughput of 74.2 img{$\cdot$}s\textsuperscript{-1} on average per subset (13.5ms per inference).

If we compare the performance scalability of each implementation, we observe an almost ideal scaling when increasing the number of active VPU chips. \autoref{fig:benchmarks_scaling} illustrates the relative performance scaling during inference by varying the batch input size on the CPU, GPU, and multi-VPU configurations. The figure reflects how well each implementation scales independently. Hence, the values are normalized per device type using their respective single-input test as reference for the normalization (i.e., 26.0ms for the CPU, 25.9ms for the GPU, and 100.7ms for the VPU). We use only one of the subsets of 10000 images from the validation dataset. In this case, we determine that the execution time required per inference is approximately reduced 50\% when duplicating the number of active VPU chips, reaching a performance increase factor of close to 8$\times$ for the last case. This matches the number of NCS devices. Nonetheless, a small penalty is observed due to the thread-management overhead and the data transferring involved. On the other hand, the performance of the CPU implementation is barely affected, with an improvement of only 14.7\% for the last case (1.1$\times$). Similar results are observed for the GPU implementation, which improves only 92.5\% for the last case (1.9$\times$). Thus, both implementations reflect relatively poor scaling in comparison.

\begin{figure*}
    \centering
    \begin{subfigure}[t]{0.4921\textwidth}
        \centering
        \hspace{-0.3921cm}
        \begin{tikzpicture}
            \begin{axis}[
                xlabel=Batch Input Size,
                ylabel=Throughtput (images/W),
                symbolic x coords={1,2,4,8},
                xtick=data,
                ytick={0,1.0,2.0,3.0,4.0,5.0},
                ymin=0,
                ymax=5.0,
                scaled y ticks = false,
                y tick label style={/pgf/number format/fixed, /pgf/number format/1000 sep = },
                ylabel style={at={(0.081,0.521)}, style={font=\small}},
                xlabel style={at={(0.5,-0.05)}, style={font=\small}},
                enlarge x limits=0.21,
                grid=major,
                legend style={legend columns=-1,at={(1.0,1.321)},anchor=north east},
                legend style={/tikz/every even column/.append style={column sep=4.21}},
                legend cell align=right,
                legend plot pos=right,
                legend style={draw=none, fill=white, inner xsep=0, inner ysep=0},
                ybar=0pt,
                bar width=5.81pt,
                width=0.81\textwidth,
                height=3.8921cm,
                area legend
            ]
            
            
            \addplot [fill=black!9] plot[error bars/.cd, y dir=both, y explicit] table [y error minus=min, y error plus=max] {
                x       y               min             max
                1	0.480614399	0.0212207741	0.0212207741
                2	0.5074685259	0.0191154729	0.0191154729
                4	0.5372794056	0.0157807751	0.0157807751
                8	0.5514572782	0.0127810095	0.0127810095
            }; \addlegendentry{CPU}
            
            \addplot [fill=black!13, postaction={pattern=crosshatch dots}] plot[error bars/.cd, y dir=both, y explicit] table [y error minus=min, y error plus=max] {
                x       y               min             max
                1	0.4818220119	0.0215615118	0.0215615118
                2	0.6589826131	0.0422467196	0.0422467196
                4	0.8251627097	0.074885981	0.074885981
                8	0.927520884	0.1029986284	0.1029986284
            }; \addlegendentry{GPU}
            
            \addplot [fill=black!21, postaction={pattern=north east lines}] plot[error bars/.cd, y dir=both, y explicit] table [y error minus=min, y error plus=max] {
                x       y               min             max
                1	3.9719375463	0.0138020826	0.0138020826
                2	3.9360407547	0.0171628867	0.0171628867
                4	3.8856057463	0.0465218618	0.0465218618
                8	3.8608786806	0.0904292815	0.0904292815
            }; \addlegendentry{VPU (Multi)}
            \end{axis}
        \end{tikzpicture}
        \caption{\textbf{Throughput-TDP Comparison} per batch size} \label{fig:benchmarks_discussion_power}
    \end{subfigure}
    \hfill
    \begin{subfigure}[t]{0.4921\textwidth}
        \centering
        \hspace{-0.3921cm}
        \begin{tikzpicture}
            \begin{axis}[
                xlabel=Batch Input Size,
                ylabel=Throughput (images/s),
                symbolic x coords={1,2,4,8,16},
                xtick=data,
                ytick={0,30,60,90,120,150,180},
                ymin=0,
                ymax=180,
                scaled y ticks = false,
                y tick label style={/pgf/number format/fixed, /pgf/number format/1000 sep = },
                ylabel style={at={(0.021,0.521)}, style={font=\small}},
                xlabel style={at={(0.5,-0.05)}, style={font=\small}},
                grid=major,
                legend style={legend columns=-1,at={(1.0,1.281)},anchor=north east},
                legend style={/tikz/every even column/.append style={column sep=4.21}},
                legend cell align=right,
                legend plot pos=right,
                legend style={draw=none, fill=white, inner xsep=0, inner ysep=0},
                bar width=9pt,
                width=0.81\columnwidth,
                height=3.96cm,
            ]
            
            \addplot [color=black, mark=*, mark options={scale=0.54}] coordinates {
                (1, 38.4491519193663)
                (2, 40.597482075146)
                (4, 42.9823524454256)
                (8, 44.1165822596966)
                (16, 44.5294257611884)
            };\addlegendentry{CPU}
            \addplot [color=black, mark=x, mark options={solid, scale=1.0921}] coordinates {
                (1, 38.5442752307935)
                (2, 52.718609046914)
                (4, 66.0130167767131)
                (8, 74.2016707173977)
                (16, 79.952636857452)
            };\addlegendentry{GPU}
            \addplot [color=black, mark=triangle*, mark options={solid, scale=0.81}] coordinates {
                (1, 9.92984386571681)
                (2, 19.6802037735531)
                (4, 38.8560574631)
                (8, 77.2175736117)
            };\addlegendentry{VPU (Multi)}
            \addplot [color=black, mark=triangle*, dashed, mark options={solid, scale=0.81}] coordinates {
                (8, 77.2175736117)
                (16, 153.0394237944)
            };
            \end{axis}
        \end{tikzpicture}
        \caption{\textbf{Projected Inference Performance} per batch size} \label{fig:benchmarks_discussion_scaling}
    \end{subfigure}
    \caption{(\subref*{fig:benchmarks_discussion_power}) Throughput performance comparison per Watt using the CPU, GPU, and multi-VPU configurations. (\subref*{fig:benchmarks_discussion_scaling}) Inference performance per batch size on a subset of the ILSVRC 2012 Validation dataset using the CPU, GPU, and multi-VPU configurations. The dashed lines represent the projected value of the multi-VPU configuration if the scaling continues. Note that, in both figures, the number of active VPU chips is coupled with the input size (1-16). \label{fig:benchmark_discussion}}
\end{figure*}

\subsection{Error Rate Comparison}

Using the same Validation dataset from the ILSVRC 2012 challenge, we evaluate the confidence accuracy of each implementation. Our goal is to understand how the differences in floating point precision can affect the predictions from the pre-trained BAIR GoogLeNet network model.

We estimate the miss-prediction rate by extracting the labels from the \emph{Validation Bounding Box Annotations} dataset of the ILSVRC 2012 challenge. For the inference error rate, we use a \emph{top-1 estimation}, as Szegedy et al.~\cite{szegedy2015going} describe. This estimation implies to accept only those predictions whose correct label has the highest confidence. In addition, we compare the absolute confidence difference between the CPU implementation\footnote{Even though the GPU implementation is excluded from the comparison, we confirm that it provides equivalent confidence results.}, which uses FP32 precision, and the VPU implementation, which uses FP16. This value reflects the average error after filtering the incorrect predictions according to the top-1 estimation.

With subtle inference error differences, we observe that the use of FP16 arithmetic on the Myriad 2 VPU does not have a major impact in the overall miss-prediction rate. \autoref{fig:benchmarks_top1} illustrates the top-1 inference error per subset using the CPU and VPU implementations. Once again, we use the validation dataset with 50000 images, and divide it in groups of 10000 subsets. From the figure, we estimate that the top-1 inference error is 31.92\% on average using the VPU. Surprisingly, the reference CPU implementation features a slightly worse error of 32.01\%. As a result, given that the top-1 inference error using the VPU implementation with FP16 arithmetic only varies 0.09\% in comparison, we confirm negligible differences due to arithmetic precision.

Looking at the absolute confidence error, we estimate once again that the use of FP16 arithmetic on the Myriad 2 VPU does not considerably affect the network output. \autoref{fig:benchmarks_diff} depicts the absolute confidence difference per subset using the VPU implementation in comparison with the CPU implementation, after filtering the top-1 miss-predictions. In this case, the average difference per subset is estimated at 0.44\% on average.

\section{Discussion}
\label{4_Discussion}



The previous results indicate that the use of VPUs can be beneficial for certain operations, such as tensor processing. Even though we observe that the execution time per inference using one chip is 4$\times$ slower compared to a reference CPU / GPU implementation, we demonstrate equivalent performance results by using a parallel, multi-VPU configuration with eight NCS devices. Yet, we have not accounted for the power consumption required on each case. In fact, the estimated thermal-design power (TDP) for both the Intel Xeon E5-2609v2 and the NVIDIA Quadro K4000 GPU used in our experiments is 80W. In comparison, the TDP of the Myriad 2 VPU is 0.9W, with an overall estimated peak consumption of 2.5W for the NCS device~\cite{anandtech2017ncs, pena2017benchmarking}. If we assume that the maximum power consumption was required\footnote{Technically, the CPU and other components are necessary to connect to the NCS (e.g., USB controller), which are not included in the estimation. Here, we only account for the operational TDP of each device.}, we can estimate a throughput function per Watt based on the number of inferences conducted per second:

\begin{equation}
{Throughput}_{Watt}=\frac{Images \cdot Second^{-1}}{TDP}
\end{equation}

By following this metric, we confirm that, in theory, VPUs could provide a throughput per Watt of over 3$\times$ higher in comparison. \autoref{fig:benchmarks_discussion_power} reflects the performance measured as images per Watt (img{$\cdot$}W\textsuperscript{-1}) for the CPU, GPU and multi-VPU configurations. From this figure, we observe that the throughput is 3.97 img{$\cdot$}W\textsuperscript{-1} when using one VPU. Increasing the number of simultaneous VPU chips does not largely affect this ratio, except for a small performance penalty due to the required data transfers. The CPU features a theoretical throughput of 0.55 img{$\cdot$}W\textsuperscript{-1} in the last case. The GPU shows similar results, with 0.93 img{$\cdot$}W\textsuperscript{-1}. Nonetheless, actual power measurements would be required in future work to understand the practical differences (i.e., the TDP can be far from the real power draws per device).


On the other hand, if we assume that the ideal scaling is maintained as we increase the number of VPU chips, we could obtain power and thermal dissipation benefits while still improving the average execution time required per inference. \autoref{fig:benchmarks_discussion_scaling} reflects this comparison using the CPU, GPU, and multi-VPU configurations. We vary the batch size from 1 to 16 inputs on the CPU and GPU implementations. In the case of the multi-VPU, we show the projected execution time after the number of NCS devices available is exceeded (i.e., eight devices). From this figure, we determine that the CPU and GPU implementations do not illustrate relevant performance improvements, with a maximum of 44.5 img{$\cdot$}s\textsuperscript{-1} and 79.9 img{$\cdot$}s\textsuperscript{-1}, respectively. The Myriad 2 VPU, however, has a projected throughput of 153.0 img{$\cdot$}s\textsuperscript{-1} using 16 VPU chips. This is a factor of 3.4$\times$ improvement over the CPU implementation, and a factor of 1.9$\times$ over the GPU version.


Despite these positive observations, we still consider that VPUs should complement the utilization of more powerful CPU and GPU architectures. For instance, it has been largely demonstrated that GPUs can be ideal for deep learning~\cite{raina2009large, coates2011analysis}. Moreover, recent architectures, such as the NVIDIA Volta V100~\cite{nvidia2017volta} or the Intel Nervana Neural Network Processor~\cite{intel2017nervana}, have been specifically designed for training and inference. 
Consequently, we foresee the potential of integrating the high-performance vector architecture featured on the Myriad 2 VPU in the form-factor of a co-processor to reduce the overall power consumption of future HPC clusters.  
Energy consumption is considered one of the main limiting factors towards the exascale supercomputer~\cite{dongarra2011international}, as we have previously motivated. In such case, 
one or several of these co-processors could be included on each node. Scientific applications could then use the VPU chips to offload certain operations that involve tensor computation, avoiding the utilization of the CPU (or GPU) on less-critical tasks. We consider to explore this path in the future. 



\section{Related Work}
\label{5_RelatedWork}


The adoption of power-efficient co-processors for computer vision and machine learning on the ``edge'' has been widely studied for robotics and the Internet-of-Things (IoT). For instance, Georgiev et al.~\cite{georgiev2014dsp} present an integrated sensing system that uses low-power DSP co-processors of commodity mobile devices to perform complex audio inferences. More specifically, Dexmont et al.~\cite{pena2017benchmarking} conduct a study of the Myriad 2 VPU for low-power robotics applications.

In the context of HPC, the use of co-processors is also frequent, specially with the emergence of the ``CPU + Accelerator'' model in this field~\cite{nickolls2010gpu}. Byun et al.~\cite{byun2017benchmarking} study the Intel Xeon Phi architecture~\cite{chrysos2014intel} as co-processor for machine learning applications. Tan et al.~\cite{tan2016accelerating}, on the other hand, propose the use of FPGAs as co-processor to accelerate Next-Generation Sequencing (NGS) applications. Notwithstanding, we observe that the integration of low-power co-processors for computation offloading is, in most cases, not considered.

Lastly, we note that the work by Ionica et al.~\cite{ionica2015movidius} shares some similarities. Here, the authors provide a comprehensive overview of the Myriad 1 VPU chip for scientific computing. In this regard, an implementation of a custom DGEMM operation that uses CMX tiling is provided, and performance results in terms of Gflops and Gflops{$\cdot$}W\textsuperscript{-1} (estimated through the TDP) are illustrated as well. While their work focuses on the opportunities that the Myriad 1 VPU chip brings for general-purpose computing, we present the technical aspects of the Myriad 2 VPU and illustrate performance results during inference on convolutional networks using multiple chips. Thus, we consider both complementary.

\section{Conclusion}
\label{6_Conclusion}

The emergence of machine learning and data-centric applications on HPC poses several constraints on general-purpose processors~\cite{jang2011exploiting, yu2015imp}. As such, power consumption and thermal dissipation become major concerns.
In this work, we have provided an overview of the Vision Processing Unit (VPU) as co-processor for inference on HPC. In particular, we have explored the most relevant technical details of the Myriad 2 VPU~\cite{moloney2014myriad, barry2015always}, in the context of the Intel Neural Compute Stick (NCS) platform~\cite{intel2017ncs}. To support our experiments, we have also presented a small inference framework, named NCSw. This framework contains a parallel, multi-VPU implementation that efficiently coordinates the execution on more than one NCS device.

Using a pre-trained network model based on the GoogLeNet work by Szegedy et al.~\cite{szegedy2015going}, we have observed that the performance during inference on a single VPU chip is only 4$\times$ slower in comparison with reference CPU and GPU implementations. By employing a multi-VPU configuration, however, we have demonstrated equivalent performance results. Yet, the expected thermal-design power (TDP) can still be reduced by a factor of 8$\times$. Moreover, we have confirmed negligible confidence differences by estimating the top-1 error rate, despite requiring FP16 arithmetic precision on the VPU.


As future work, we expect to conduct a thorough study of the possibilities of the Myriad 2 VPU as co-processor for task offloading on HPC. This would imply extending our work and integrating the VPU chip as a conventional vector processor for general-purpose computing. In addition, we expect to compare the VPU with highly-specialized accelerator chips, such as the NVIDIA Volta V100 architecture~\cite{nvidia2017volta}. This would give us a better understanding of the benefits in contrast with recent, novel architectures designed for machine learning.

 \section*{Acknowledgements}
The experimental results were performed on resources provided by the Swedish National Infrastructure for Computing (SNIC) at PDC Centre for High Performance Computing (PDC-HPC).

The work was funded by the European Commission through the SAGE project (Grant agreement no. 671500 / http://www.sagestorage.eu). 


\bibliography{main}


\end{document}